# The LCG POOL Project – General Overview and Project Structure

Dirk Duellmann on behalf of the POOL Project
*European Laboratory for Particle Physics (CERN), Genève, Switzerland*

The POOL project has been created to implement a common persistency framework for the LHC Computing Grid (LCG) application area. POOL is tasked to store experiment data and meta data in the multi Petabyte area in a distributed and grid enabled way. First production use of new framework is expected for summer 2003. The project follows a hybrid approach combining C++ Object streaming technology such as ROOT I/O for the bulk data with a transactionally safe relational database (RDBMS) store such as MySQL. POOL is based a strict component approach - as laid down in the LCG persistency and blue print RTAG documents - providing navigational access to distributed data without exposing details of the particular storage technology. This contribution describes the project breakdown into work packages, the high level interaction between the main pool components and summarizes current status and plans.

## 1. INTRODUCTION

Data processing at LHC[1] will impose significant challenges on the computing of all LHC experiments. The very large volume of data – some hundred Petabytes over the lifetime of the experiments – requires that traditional approaches, based on explicit file handling by the end user, be reviewed. Furthermore the long LHC project lifetime results in an increased focus on maintainability and change management for the experiment computing models and core software such as data handling. It has to be expected that during LHC project lifetime several major technology changes will take place and experiment data handling systems will be required to be able to adapt quickly to the changes in the environment or the physics research focus.

In the context of the LHC Computing Grid (LCG[2]) a common effort to implement a persistency framework underlying the different experiment frameworks has been started in April 2002. The project POOL[3] (acronym for **P**OOL **O**f persistent **O**bjects for **L**HC) has since then ramped up to about 10 FTE from IT/DB group at CERN and the experiments located at CERN and outside institutes.

For POOL as a project, the strong involvement of the experiments from the very early stages on is seen as very important to guarantee that the experiments' requirements are injected and implemented by the project without introducing too much distance between software providers and users. Many of the POOL developers are part of an experiment software team and will be directly involved also the integration of POOL into their experiments software framework.

### 1.1. Component Architecture

POOL as a LCG Application Area project follows closely the overall component base architecture laid down in the LCG Blueprint RTAG report[4]. The aim is to follow as much as possible a technology neutral approach. POOL therefore provides a set of service APIs - often via abstract component interfaces - and isolates experiment framework user code from details of a particular implementation technology. As a result, the POOL user code is not dependent on implementation API or header files, POOL applications do not directly depend on implementation libraries. Even though POOL implements object streaming via ROOT-I/O[10] and uses MySQL[11] as an implementation for relational database services, there is no link time dependency on the ROOT or MySQL libraries. Back end component implementations are instead loaded at runtime via the SEAL[5] plug-in infrastructure. The main advantage of this approach is that changes required to adapt to new back end implementations are largely contained inside the POOL project rather than affect the much larger code base of the experiment frameworks or even end user code. Achieving this goal and still keeping the system open for new developments is only possible by constraining very consciously the concepts exposed by POOL. The project has made a significant effort to identify a minimal API that is just sufficient to implement the data management requirements but still can be implemented using most implementation technologies which are available today.

### 1.2. Hybrid Technology Store

The POOL system is based on a hybrid technology approach. POOL combines two main technologies with quite different features into a single consistent API and storage system. The first technology includes so-called object streaming packages (eg ROOT I/O) which deal with persistency for complex C++ objects such as event data components. Often this data is used in a write-once, read-many mode and concurrent access to the data can therefore be constrained to the simple read-only case. This simplifies in particular the deployment as no central services to implement transaction or locking mechanisms are required. The second technology class provides Relational Database (RDBMS) services such as distributed, transactionally consistent, concurrent access to data which still can be updated. RDBMS based stores also provide facilities for efficient server side query evaluation. The aim of this hybrid approach is to allow users to be able to choose the most suitable storage implementation in for different data types, use cases and deployment environments. In particular RDBMS based components are currently used heavily in the area of catalogs, collections and their meta data, streaming technology is used for the bulk data.

### 1.3. Navigational Access

POOL implements a distributed store with full support for navigation between individual data objects. References between objects are transparently resolved – meaning that referred-to objects are brought into the application memory automatically by POOL as required by the application. References may connect objects in either the same file or spanning file and even technology boundaries. Physical details such as file names, host names and the technology which holds a particular object are not exposed to reading user code. These parameters can therefore easily be changed which allows optimizing the computing fabric with minimal impact on existing applications.

## 2. PROJECT BREAKDOWN INTO WORK PACKAGES

The internal structure of POOL follows closely a domain decomposition which has been described to a large extend already in the report of the Persistency RTAG[7] which preceded the POOL project. In this paper we give only a brief overview on the overall project structure and the main responsibilities and collaboration between its main components. A more detailed description of component implementations can be found in [8] and [9]. Component design documents are available at [3].

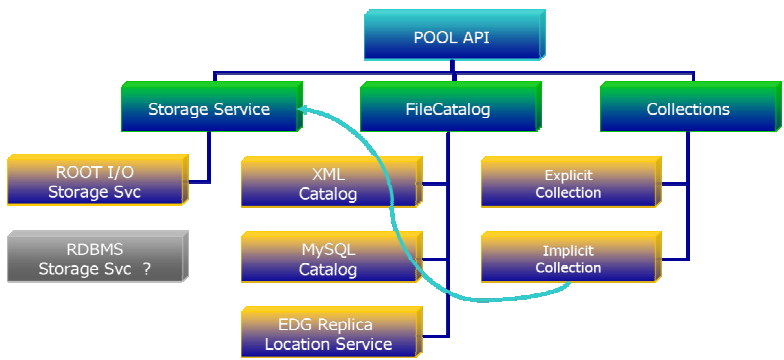

Fig. 1, POOL breakdown in components

### 2.1. POOL Storage Hierarchy

The storage hierarchy exposed by POOL consists of several layers (shown in Fig. 2) each dealing with more granular objects than the one above. The entry point into the system is the POOL context which holds all objects which have been obtained so far. Each context may reference objects from any entry in a given File Catalog. Currently POOL supports a single File Catalog at a time – this may be extended in later releases. By specifying the file catalog for a particular application one determines the scope of objects this application can see.

The context is also the granularity of user level transactions which are provided by POOL. All objects in a context which have been marked for writing will be written together at the context transaction commit. The persistency service subcomponent of the storage service keeps a list of open database connections and issues individual low level commits on the database level as required.

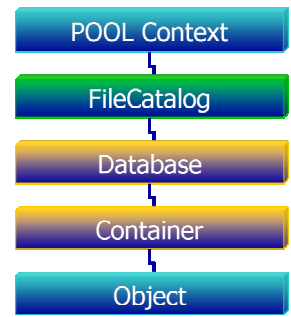

Fig.2 POOL Storage Hierarchy

Each POOL database (entry in the POOL file catalog) has a well defined major *storage technology*. Currently only one major technology is supported - namely ROOT I/O files - are supported but the RDBMS storage manager prototype will be a first extension to prove that such independence has indeed been achieved.

POOL databases are internally structured into containers which are used to group persistent objects inside the database. POOL containers in the same database may differ in their minor technology type but not in their major type (eg a single ROOT I/O database file may hold containers of ROOT-tree and ROOT-keyed type).

Some storage service implementations may constrain the choice of data types which can be kept in a container simultaneously. For example a ROOT tree based container does not allow storing arbitrary combinations of unrelated types in the same container, a ROOT directory based container does.

### 2.2. File Catalog

The main responsibility of the File Catalog is to keep track of all POOL databases (usually files which store objects) and to resolve file references into physical file names which are then used by lower level components like the storage service to access file contents. More recently the POOL file catalog has been extended to allow simple meta data to be attached to each file entry. This infrastructure is shared with the collection implementation.

When working in a Grid environment a File Catalog component based on the EDG Replica Location Service (RLS) is provided to make POOL applications grid aware. File resolution and catalog meta data queries in this case are forwarded to grid middleware requests.

For grid-disconnected environment MySQL- and XML-based implementations of the component interface exist, which use a dedicated database server in the local area network (eg isolated production catalog servers) or local file system files (eg disconnected laptop use cases).

Files are referred to inside POOL via a unique and immutable file identifier (FileID) which is assigned at file creation time. This concept of a system generated FileID has been added by POOL to the standard grid model of many-to-many mapping between logical and physical file names to provide for stable inter-file reference in an environment were both logical and physical file names may change after data has been written. The stable FileID allows POOL to maintain referential consistency between several files which contain related objects without requiring any data update eg to fix up changes in logical or physical file names.

In addition the particular FileID implementation which has been chosen for POOL which is based on so-called Universally or Globally Unique Identifiers (UUID/GUID[12]) provides another very interesting benefit. GUID based unique FileIDs can be generated in complete isolation without a central allocation service. This greatly simplifies the distributed deployment of POOL, as POOL files can be created even without network connection and still later be integrated in a much larger store catalogs without any risk of clashes.

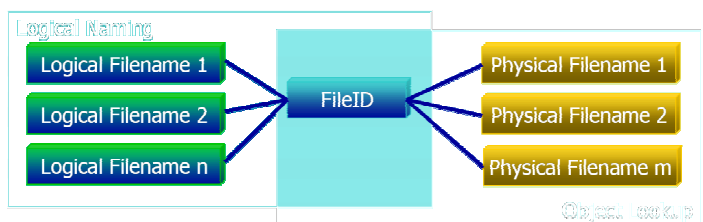

Fig 3. POOL File Catalog Mapping

## 2.3. Storage Service & Conversio

The storage technology information from the File Catalog is used to dispatch any read or write operation to a particular storage manager. The task of the storage manager component is to translate (stream) any transient user object into a persistent storage representation which is suitable for reconstructing an object in the same state later. The complex task of mapping individual object data members and the concrete type of the object relies on the LCG Object Dictionary component developed by the SEAL project. For each persistent class this dictionary provides detailed information about internal data layout which is then used by the storage service to configure the particular backend technology (eg ROOT I/O) to perform I/O operations.

In addition to the existing storage service which supports objects in ROOT trees and objects in ROOT directories, a prototype implementation of a RDBMS base store is underway. As the POOL program interface hides the details of their internal implementation, the user can easily adapt to new requirements or technologies with very little change to the application code.

During the process of writing an object a unique object identifier is defined, which can later be used to locate the object throughout a POOL store.

## 2.4. Object Cache & References

Once an object has been created in application memory, either by the user to be written out or as a result of a POOL read operation, the object is maintained in an object cache (also called Data Service) to speed up repeated accesses to the same object and control object lifetime. The implementation provided with POOL uses a templated smart pointer type (pool::Ref<T>) which implements – close to the ODMG standard - object loading on demand and automatic cache management via reference counting on all cached object.

Alternatively an experiment may decide to clean all objects from the cache via an API explicitly or to replace the POOL object cache with its own implementation by providing an implementation of the cache interface defined in POOL.

As the inter-object references can be stored part of a persistent object as well, and as POOL will transparently load objects on demand, the ref is also the main building block to construct persistent associations between objects. These may be local to a single file but also across file and even technology boundaries. Object lifetime management and object caching are coupled closely to the user implementation language – currently C++ for LHC offline code. This POOL component therefore to a large extend acts as a C++ binding of POOL and encapsulates most functional changes which would be required in case native support additional language should become a requirement.

## 2.5. Collections

The collection support provided by POOL allows maintaining large scale object collections (eg event collections) – and should not be confused with the standard C++ container support which is provided by the POOL storage service. POOL collections can be optionally extended with meta data (currently only simple lists of attribute-value pairs) to support user queries to select only collection elements which fulfill a query expression. POOL supports several different collection implementations based either on the RDBMS back end or on the streaming layer. Collections can be defined explicitly – via adding each contained objects explicitly – or as so-called implicit collection which refers to all objects in a given list of databases or containers. As the different collection implementations adhere to a common collection component interface, the user can easily switch from a collection using ROOT trees in local files to database implementations which allows distributed access and server side query evaluation.

## 3. PROJECT ORGANISATION

The POOL internal project structure is closely aligned with the functional decomposition of the system. Three work packages have been created to implement the Storage, File Catalog and Collection component services discussed above. A fourth work package deals with release coordination, testing and the overall POOL development infrastructure relying on core services provided by the LCG-SPI[6] project.

### 3.1. Release Procedure

POOL follows the rapid release cycles proposed by the LCG software process RTAG[13] with the aim to stimulate early feedback from the participating experiments. Roughly once every 6 weeks a public release is produced and announced to the user community. Internal releases which are publicly available but without complete testing, component set and documentation are produced several times a week. As the POOL developers are distributed and - because of the componentization – are usually exposed to a subset of the POOL functionality only, we organize weekly work package meetings and a bi-weekly full project meeting. Roughly every 2 release cycles we perform an internal code review to increase the common knowledge about component implementations in other project areas.

### 3.2. Status and Plans

At the time of writing the project is finishing POOL V1.1 which will add significant functional enhancements on the request of the experiments (in place update on the streaming layer, support for more STL containers, transient data members, new simplified transaction model etc.). So far the project has fulfilled its milestones without significant delays and has shown that it can cope with requirement changes and new feature requests rapidly. Both the strict decoupling between different component implementations and the largely automated testing framework setup in collaboration with the LCG SPI project have significantly contributed to that success.

Given the short project duration POOL is still a relatively young package and a real prove that all relevant requirements have been identified and are met will need the successful completion of a significant pre-production activity (eg as part of an experiment production effort like the upcoming CMS pre-production). Undoubtedly this will reveal some remaining implementation problems and help to prioritize the remaining developments of POOL. Current items on the POOL work list include the introduction of an RDBMS vendor independence layer for relational components, a proof-of-concept prototype on an RDBMS based storage service and

## 4. SUMMARY

The LCG POOL project provides a new persistency framework implemented as a hybrid data store. It integrates seamlessly existing streaming technology (eg ROOT I/O) for complex object storage with RDBMS technology (eg MySQL) for consistent meta data handling. Strong emphasis has been put on strict component decoupling and well defined inter component communication and dependencies.

POOL provides transparent cross-file and cross-technology object navigation via C++ smart pointers without requiring the user to explicitly open individual files or database connections. At the same time it is via the EDG-RLS based catalogue integrated with Grid technology. The component architecture preserves the possibility to choose at runtime between networked and grid-decoupled working modes

The recently produced POOL V1.0 release is currently integrated in several of the experiment frameworks and is expected to be first deployed in production activities this summer. Essential for the success will be a tight connection to experiment development and production teams to validate the feature set and tight integration with LCG deployment activities.

### References

[1] LHC – The Large Hadron Collider,
   http://www.cern.ch/lhc
[2] The LHC Computing Grid
   http://lcg.web.cern.ch
[3] The POOL Project,
   http://pool.cern.ch
[4] T. Wenaus *et al*, Report of the LHC Computing Grid Project Architecture Blueprint RTAG
   http://lcgapp.cern.ch/project/blueprint/BlueprintReport-final.doc
[5] J. Generowicz *et al.*, SEAL: Common Core Libraries and Services for LHC Applications,
   CHEP 2003 proceedings, MOJT003.
   see also: http://seal.web.cern.ch/seal
[6] The LGC SPI Project.
   http://spi.cern.ch
[7] D. Malon *et al.*, Report of the LHC Computing Grid Project Persistency Management RTAG,
   http://lhcgrid.web.cern.ch/LHCgrid/sc2/RTAG1/
[8] M.Frank *et al.*, The POOL Data Storage, Cache and Conversion Mechanism.
   CHEP 2003, proceeding, MOKT008
[9] Z.Xie *at al.*, POOL File Catalog, Collection and Meta Data Components,
   CHEP 2003, proceeding, MOKT009
[10] R.Brun and F.Rademakers,
   ROOT-An Object Oriented Data Analysis Framework,
   Nucl. Inst.&Meth. in Phys.Res.A389(1997)81-86.
   see also: http://root.cern.ch
[11] MySQL – Open Source Database,
   http://www.mysql.com/
[12] P. Leach, R. Salz,'UUIDs and GUIDs', Internet-Draft,

ftp://ftp.isi.edu/internet-drafts/draft-leach-uuids-guids-00.txt
February 24, 1997.
[13] F. Rademakers et al, Report of the LHC Computing Grid Project Software Management Process RTAG
http://lhcgrid.web.cern.ch/LHCgrid/SC2/RTAG2/finalreport.doc
[14] EDG Replica Location Service,
http://cern.ch/edg-wp2/

**MOKT007**